\documentclass[printer]{aa}
\usepackage{graphicx}
\usepackage{txfonts}
\usepackage{natbib}
\usepackage{longtable}
\bibpunct{(}{)}{;}{a}{}{,}

\begin{document}
\title{On the Fundamental Plane of the Galactic Globular
Cluster System}
\author{M. Pasquato \inst{1}
\and G. Bertin \inst{2}} \institute{Dipartimento di Fisica,
Universit\`a di Pisa, Largo Bruno Pontecorvo 3, I-56127 Pisa, Italy
\and Dipartimento di Fisica, Universit\`a degli Studi di Milano,
via Celoria 16, I-20133 Milano, Italy}
\date{Received 27/01/2008 / Accepted 09/05/2008}
\authorrunning{Pasquato \& Bertin}
\titlerunning{Fundamental Plane of Globular Clusters}

\abstract{The globular clusters of our Galaxy have been found to
lie close to a plane in the $\log{R_e}$, $\log{\sigma}$, ${\mathit{SB}}_e$
space (see Djorgovski 1995), on the continuation of the
Fundamental Plane that is known to characterize the global
properties of early-type galaxies. Of course, there is no apparent
reason why such physically different self-gravitating systems should
follow the same scaling law.}{We reexamine the issue by focusing
on a sample of 48 globular clusters selected in terms of
homogeneity criteria for the photometric data available from the
literature.}{We perform a model-independent analysis of surface
brightness profiles and distance moduli, estimating error bars and
studying selection effects with robust non-parametric
statistical tests.}{We determine the values of the coefficients
that define the Fundamental Plane and their error bars and show
that the scatter from the Fundamental Plane relation is likely to
be intrinsic, i.e. not due to measurement errors only.
Curiously, we find that in the standard Fundamental Plane
coordinates the set of points for our sample occupies a rather
slim, axisymmetric, cylindrical region of parameter space, which suggests that
the relevant scaling relation might be around a line, rather than a plane,
confirming a result noted earlier. This is likely to be the origin
of the difficulties in the fit by a plane often mentioned in
previous investigations. In addition, such a Fundamental Line relation
would imply a pure photometric scaling law relating luminosity to the
effective radius which might be tested on wider samples and on
extra-galactic globular cluster systems. As to the residuals from
the Fundamental Plane relation, we find a correlation of the deviations
from the plane with the central slope of the surface brightness profile.
No other statistically significant correlations are identified.
Finally, given the constraint imposed by the virial theorem, we
study the distribution of the values of the quantity $K_V/(M/L)$
(virial coefficient divided by the relevant mass-to-light ratio);
the distribution of the logarithms, reconstructed through kernel
density estimation methods, shows evidence for bimodality, which
suggests that the galactic globular cluster system may be composed
of at least two dynamically different populations. Yet, these
populations do not appear to reflect the standard dichotomy
between disk and halo clusters.}{} 

\maketitle

\section{Introduction}

It is well known \citep{1987ApJ...313...42D, 1987ApJ...313...59D}
that early-type galaxies occupy a region in the vicinity of the
plane $\log{R_e} = \alpha \log{\sigma_0} + \beta \mathit{SB}_e + \gamma$ in
the three-dimensional parameter space defined in terms of the
projected effective radius\footnote{Some authors employ $r_e$, $R_h$, or $r_h$ to refer to the projected effective (half-light) radius. In our notation $r_e$ refers to the projected effective \emph{angular} radius, while $R_e$ to the projected effective \emph{linear} radius.} $R_e$, of the central velocity dispersion $\sigma_0$, and of the mean surface brightness $\mathit{SB}_e$. Such Fundamental Plane (FP) exhibits a tilt, i.e. the measured
values of the coefficients that characterize the plane differ
significantly from the naive expectations based on the application
of the virial theorem. The properties of the FP reflect partly the
characteristics of the stellar populations in these systems and
partly some systematic deviations from strict structural homology
\citep[e.g., see][]
{2002A&A...386..149B}.

It is often stated that the universality of the FP extends well
outside the domain of early-type galaxies and applies to many
other systems, such as globular clusters
\citep[hereafter GCs; see ][]{1995ApJ...438L..29D, 2000ApJ...539..618M, 2007AJ....133.2764B}, ``cluster spheroids'' \citep[][]{2006ApJ...638..725Z}, open clusters
\citep{2005A&A...437..483B}, X-ray emitting gas in elliptical
galaxies \citep{2005ApJ...633L..21D}, clusters of galaxies
\citep[see][and references therein]{2004ApJ...600..640L},
and even more exotic objects, such as supermassive black holes 
\citep[e.g.,][]{2007ApJ...669...67H} and quasars 
\citep[e.g.,][]{2006NewAR..50..758H}.
Comparisons of the FP relation for different systems are often present
in the literature: for example, see \cite{1997AJ....114.1365B} and \cite{2004ApJ...610..233M}, who also address the problem of the mass
gap between high-mass GCs and low-mass galaxies and related classification issues in the so-called $k$-space introduced by \cite{1992ApJ...399..462B}.

In particular, for galactic GCs it has been found by
\cite{1995ApJ...438L..29D} that an FP relationship based on core
parameters (core radii and central surface brightness) holds and
that it is ``consistent with globular cluster cores being
virialized systems with a universal and constant $M/L$ ratio",
while for quantities referred to the
effective radius ($R_e$ and $\mathit{SB}_e$) the FP does not emerge as
sharply, possibly because of error correlations (see also
additional remarks in Sect.~\ref{sec:FP} below).
In reality, since GCs are non-homologous stellar systems with
rather homogeneous stellar populations, it would be surprising
if they happened to follow a continuation of the FP of early-type
galaxies, which appear to be associated with very different
structural properties, in terms of homology, stellar populations,
and dark matter content.

In this paper we thus reconsider the problems of the existence and
of the nature of the FP for the GCs of the Milky Way Galaxy.
Although galactic GCs are extremely well studied objects, for a
proper statistical investigation of their structural properties
the use of the entire set of data available from the literature
suffers from several limitations and is therefore problematic. In
fact, since the GCs are basically distributed over the entire sky,
the relevant photometric and spectroscopic data that have been
collected come from different telescopes and different
instruments. In addition, GCs are located at different distances
from the Sun and, in the sky, at very different ranges of galactic
longitude and latitude, thus suffering from different amounts
of reddening. Thus different groups of GCs suffer from different
uncertainties in distance determination, which has a major impact
on the determination of the intrinsic properties that enter the
FP. All this makes the entire set of data highly inhomogeneous and
thus not well suited for a proper statistical investigation.

As a preliminary step for a satisfactory statistical
investigation, we have thus decided to identify a sample of
galactic GCs that, in our view, has optimal characteristics, being
sufficiently large while being associated with sufficiently
homogeneous data in relation to their photometric properties. Our
study of the FP has thus been restricted to such selected sample.
Sample selection is a delicate issue, so a thorough discussion of
this problem is carried out in the following.

The paper is organized as follows. In Sect.~2 we describe the
criteria that have led us to identify our optimal sample of GCs
and then summarize its properties. In Sect.~3 we proceed to study,
for such sample, the possible existence of the FP in the relevant
parameter space and show that, in the standard FP
coordinates, our sample occupies a rather slim cylindrical region,
which suggests that the scaling relation might be around a line,
rather than a plane, a result already noted in previous
investigations \citep[Bellazzini 1998, see also][]{1997AJ....114.1365B}.
In Sect.~4 we discuss two unexpected features, i.e. a residual correlation
with a parameter characterizing the central slope of the brightness
profile and a bimodality in the distribution of the effective virial
coefficients $K_V/(M/L)$, reconstructed by means of kernel density
estimation methods. In Sect.~5 we discuss the selection effects on
our sample in comparison with those present in previous
investigations \citep[in particular, in][]{1995ApJ...438L..29D}.
Finally, in Sect.~6 we summarize the results obtained in the paper
and present discussion and conclusions.

\section{The adopted sample}

Since our primary objective is to test whether indeed the galactic
GCs follow a continuation of the FP of early-type galaxies, we
have decided to refer to the standard global photometric
parameters, i.e. to the effective radius and to the related mean
surface brightness ($R_e$ and $\mathit{SB}_e$), which are commonly used in
the study of the FP of early-type galaxies (for a complete
definition, see Bertin et al. 2002) and to ignore other options
(such as the use of core parameters, $R_0$ and $\mathit{SB}_0$; see
Djorgovski 1995). The criteria for the identification of our
sample, extracted from the data set available from the literature,
are largely dictated by the requirement of accurate distance
determination.

In the following four subsections, we address separately the main
ingredients required for an accurate determination of the
parameters entering in the FP relation, i.e. distance moduli,
reddening, surface brightness profiles, and line-of-sight velocity
dispersions. We will then complete the Section by identifying our
sample.

\subsection{Distance moduli}
The two most recent papers that contain an extensive set of
galactic globular clusters for which distance moduli have been
determined with a uniform method (using the Zero Age Horizontal Branch (ZAHB) as a standard candle) and have been assigned individual error bars
are the article by \cite{1999AJ....118.1738F}, who studied and
reduced uniformly colour-magnitude diagrams coming from several
sources, for a set of 61 GCs and that by \cite{2005A&A...432..851R}, based on an
extensive photometric data set from HST for $72$ GCs \citep{2002A&A...391..945P}.
The samples overlap partially. Different procedures are used in the
two papers to deal with data reduction, particularly with respect to the
determination of the ZAHB level. \cite{2005A&A...432..851R} compare their
distance moduli whith those of \cite{1999AJ....118.1738F} on the
intersection of the two samples, and find them on average larger than
those of \cite{1999AJ....118.1738F} with a mean offset of $0.09$ mag.
Moreover, an rms scatter of $0.17$ mag is found for the differences,
with some values being as discrepant as $0.4$ mag.
Therefore, in this paper, we decided to consider the properties of a
combined sample, which contains $48$ GCs (see Sect. \ref{sect:2.5}), but also to check the behaviour of two sub-samples studied separately (see Appendix).

In their Table 2, \cite{1999AJ....118.1738F} list ZAHB
magnitude levels, with the related photometric error bars, and two
sets of true distance moduli (${|m - M|}_0$), based on two
different assumptions on the enhanced abundance of
$\alpha$-elements (the so-called $[\alpha/Fe]$ enhancement) of GC
stars with respect to solar abundances. Column $7$ of that Table
lists the distance moduli obtained by assuming solar abundances
(i.e., with no $[\alpha/Fe]$ enhancement) and $[Fe/H]$ from
\cite{1997A&AS..121...95C}, while column $8$ lists the distance
moduli obtained by adopting $[\alpha/Fe]$ values from the
literature \citep{1996A&A...305..858S,
1996PASP..108..900C} and by mimicking $\alpha$-enhanced
isochrones by standard scaled solar models of suitable
metallicity, as proposed by \cite{1993ApJ...414..580S}. The
authors state that distance moduli ``are affected by many
uncertainties (namely, the evaluation of the ZAHB level, the zero
point and dependence on metallicity of the ZAHB level, reddening,
etc.)" so that ``the global uncertainty affecting the distance
moduli listed in Table 2 cannot be less than $0.2$ mag".

If we linearly interpolate the distance moduli as a function of
$[\alpha/Fe]$ on the interval between $0$ and the value of
$[\alpha/Fe]$ adopted for the construction of column $8$, we can
take into account the uncertainty on $[\alpha/Fe]$ by adding to
the distance moduli error the product between the relative error
in $[\alpha/Fe]$ and the difference between the distance moduli
listed in Table $8$ and those in Table $7$. This appears to be
reasonable. In view of the sparseness and heterogeneity of
$[\alpha/Fe]$ data, as noted also by \cite{1999AJ....118.1738F},
we estimate the relative error on this quantity to be
approximately $50\%$.
With this device, we can take into account this factor in the
uncertainty on distance moduli. Keeping in mind the level of $0.2$
mag suggested by \cite{1999AJ....118.1738F}, we decided to proceed
by linearly transforming the error bars obtained above, so that
the first quartile of the error bar size distribution was rescaled
into $0.2$ mag and the third quartile into $0.3$ mag.
This procedure is both reasonable and statistically robust,
because it is not affected by outliers in the error bar size
distribution and turns out to be compatible with the scatter and
systematic differences that arise when comparing distances
obtained from different standard candles, such as the Main
Sequence or the White Dwarf Cooling Sequence \citep[for a distance
measurement of NGC $104$ based on the White Dwarf Cooling
Sequence, see][]{1999ASPC..167..336Z}, or even when comparing the distance moduli of \cite{1999AJ....118.1738F} with those of
\cite{2005A&A...432..851R}.

The distance moduli of \cite{2005A&A...432..851R} are probably of higher quality
with respect to those of \cite{1999AJ....118.1738F}, because they are based on
uniform HST photometry, at least as far as random error bars
are concerned.
For the intersection of the two samples our adopted uncertainties are therefore
obtained by summing in quadrature to the error estimates by \cite{2005A&A...432..851R} the semi-difference between the distance moduli obtained by \cite{2005A&A...432..851R} and those by \cite{1999AJ....118.1738F}.
For the remaining clusters with distance moduli only in \cite{2005A&A...432..851R}, we assume a constant difference equal to the mean difference quoted above, and use it to do an analogous quadrature sum to get the adopted error bars.

Our conservative procedure may tend to overestimate the error bars
on distance moduli. Still we will see that the formal reduced
$\chi^2$ for the FP fit will turn out to be larger than
unity.

\subsection{Reddening}
\cite{2005A&A...432..851R} provide independent measurements of the
reddening affecting the GCs in their sample, which we averaged with those
reported in the McMaster Catalogue \citep{1996AJ....112.1487H}.
For the GCs not considered by \cite{2005A&A...432..851R}, we just adopted the values reported in the latter Catalogue.
On the values of reddening, we assigned error bars as suggested by Harris: ``The
typical uncertainty in the reddening for any cluster is on the order
of $10$ percent, i.e. $\delta[E(B-V)] = 0.1 E(B-V)$".
Note that the apparent distance moduli determined by \cite{2005A&A...432..851R}
refer to the F555W HST band; to de-redden them, we thus
applied the relevant reddening/extinction relationship, obtained from Table 12 of \cite{1995PASP..107.1065H} through a linear fit of the form $A_{F555W} = k E(B - V)$ and differs from
the usual $A_V = 3.1 E(B - V)$ essentially in the value of the proportionality
coefficient $k$.

\subsection{Surface brightness profiles: apparent
magnitudes and effective radii}

To calculate apparent effective radii $r_e$ and apparent V-band magnitudes (or equivalently, the values of mean V-band surface brightness within the effective radius, $\mathit{SB}_e$) we started from the circularized surface brightness profiles reported by \cite{1995AJ....109..218T}. A more recent compilation \citep{2005ApJS..161..304M} of GC surface brightness profiles and structural parameters is available, comprising both Milky Way and extragalactic GCs, but, for the galactic GCs, it is based on fitting various models to the same set of data by \cite{1995AJ....109..218T}. Our model independent approach thus prompted us to proceed independently, basing our analysis directly on the \cite{1995AJ....109..218T} profiles. These profiles are based on data from different sources and do not quote explicit error estimates. On the other hand, the authors provide some quality rating of the data, based on the reliability of the data-set from which they were extracted. These weights, $w$, ranging from $0.0$ (totally unreliable data-set) to $1.0$ (high
quality data-set) were admittedly ``assigned by eye" (Trager, private communication). Relative error bars might be introduced as $\sigma \propto 1/\sqrt{w}$, but this would still be somewhat arbitrary. Eventually, to set the magnitude of the error bar for each surface brightness data-point, we resorted to the (halved) residuals to the Chebyshev polynomial fits to the profiles provided by \cite{1995AJ....109..218T}, because we wished to avoid relying on a subjective assessment of data quality. This choice has its own limitations, because such residuals depend on the particular form of the fitting function, are partly correlated, and may include as an error term some
spurious deviations (e.g., deviations associated with the presence
of a bright star).

The physical parameters that we wish to derive require integration
of the profiles on the radial coordinate. Since the points are not
evenly spaced in radius and are rather noisy, we have decided to
smooth and interpolate the data by cubic splines, prior to
integration \citep[see also][]{1995AJ....109..218T}. We then
proceeded to integrate using different methods (e.g., trapezoidal and Simpson's rule) and checked that the
results were stable. The integration error was obtained by
comparing results derived by changing the number of interpolating
points and controlled to be at least one order of magnitude
smaller than other sources of error.

The surface brightness profiles were extrapolated beyond the last
available data-point, which is set by the merging of the profile
into the background sky brightness and so differs from one GC to
another. To extrapolate, we used a linear relation between surface
brightness and angular radius. We tested a number of alternatives
for an optimal extrapolation and noted that the result did not
change the derived total magnitude by more than $0.05$ mag.

The apparent effective radii were then determined in a straightforward
manner by building a curve of growth through integration of the
profile and finding the radius at which the integrated luminosity
equals half of the total luminosity. Obviously, since uncertainty
on the total magnitude propagates to the effective radius, this
factor has the potential of introducing an anti-correlation
between uncertainties on the two quantities \citep[see also Sect. 3.1 of][]{1999MNRAS.308.1037T}. The importance of
this error term, which in the error budget for the effective
radius competes with the error deriving from the uncertainty on
the parameters of the interpolation curve, depends strongly on the
steepness of the growth curve in the vicinity of the effective
radius.

\begin{figure}
  \resizebox{\hsize}{!}{\includegraphics[angle =
270]{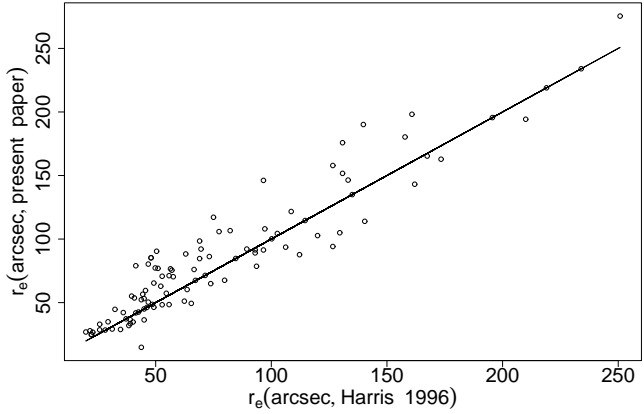}}
  \caption{Comparison of apparent effective radii reconstructed in
this paper with those reported in the McMaster Catalogue
\protect\citep{1996AJ....112.1487H}. A $1:1$ line has been drawn to guide the eye.}
  \label{myradsvsharris}
\end{figure}

Fig.~\ref{myradsvsharris} shows a comparison of the values of effective radii
obtained in this paper with those recorded in the McMaster
Catalogue \citep{1996AJ....112.1487H}. Following
\cite{1995AJ....109..218T}, we replaced points characterized by
excessive discrepancies with values from the literature \citep[in our
case from][]{1996AJ....112.1487H}. This was performed by
replacing only those data that we found to be outliers in the
distribution of differences. For this purpose, we defined as
outlier a point which lies outside the interval $[Q_1 - (Q_3 -
Q_1), Q_3 + (Q_3 - Q_1)]$, with $Q_1$ and $Q_3$ being the first
and third quartile of the data distribution; this is slightly more
conservative with respect to the more common definition based on
$[Q_1 - 1.5(Q_3 - Q_1), Q_3 + 1.5(Q_3 - Q_1)]$.

Our angular radii are slightly larger than those listed in the McMaster Catalogue, showing a median difference of $0.04$ in the logarithm. The standard deviation of the differences is somewhat larger, at $0.10$, but still compatible with our estimated error bars on effective radii. Similarly, our magnitudes exceed those of \cite{1996AJ....112.1487H} by $0.07$ mag in the median (i.e., we find clusters to be systematically dimmer), and the differences show a standard deviation of $0.21$ mag. This scatter is slightly higher than that expected based on error estimates for our sample alone.

Unfortunately, the surface brightness profiles presented by
\cite{1995AJ....109..218T} do not allow us to carry out the
analysis in different colors. Therefore, we will be unable to
address the possible impact of mass segregation on the FP scatter.

\subsection{Velocity dispersions}

We took the line-of-sight velocity dispersions from
\cite{1993ASPC...50..357P}, which are complete with uncertainties
estimated by the authors. The term \emph{central} velocity
dispersion (and the corresponding subscript $0$) would be misleading in this context; so, while we have used it when referring to the FP of elliptical galaxies, we will drop it in the following,
because the data set contains some dispersions that are not
strictly referred to the center of the cluster.
We did not try to rescale these measurements to a common aperture size,
because this would be, to some extent, model dependent (e.g., if we made use of \cite{1966AJ.....71...64K} models; see also \cite{2005ApJS..161..304M} and Table 13 therein for observed central velocity dispersions corrected by using different models). In the adopted data set two kinds of velocity dispersion measurements are mixed, from integrated light spectroscopy and from the kinematics of individual stars. On a total of $56$ GCs, $31$
have integrated light measurements and $35$ have stellar kinematic
measurements, while only $10$ have both.

Integrated light measurements make use of a small (several
arcseconds) spectroscopic slit centered on the GC. In contrast,
samples of stars used for stellar kinematics often span a much
wider angle (up to several arcminutes). In addition, integrated
light spectroscopy cannot discriminate between the cluster and
background or foreground sources, nor can it exclude fast moving
binary stars which may exaggerate the observed velocity
dispersion. Moreover, the light collected by the spectroscopic
slit may be dominated by a small number of bright giants, therefore
amounting to poor statistics. On the other hand, stellar samples used
for kinematics are usually selected carefully for membership and
tested against binaries through repeated observations, but may be
rather small and unevenly distributed in the cluster; they also
are often biased towards highly luminous stellar types (mainly red
giants). All these factors contribute to producing a systematic
difference between these two kinds of measurements, which we have
tested on the subsample of clusters having both kinds of measurement
and found to be significant.
In Sect. \ref{Sect:distkv}, devoted to the analysis of the distribution of virial coefficients, we will restrict our study to the relatively homogeneous subsample of GCs for which the velocity dispersion is given by integrated light measurements, because the virial coefficient is proportional to the square of the velocity dispersion and sample homogeneity is crucial. For the rest of the paper, we repeated our calculations on the restricted subsample and checked that the results are in general agreement with those obtained on the larger, unrestricted sample.

\subsection{The sample}
\label{sect:2.5}

The McMaster Catalogue of Milky Way Globular Cluster Parameters
\citep{1996AJ....112.1487H} lists $151$ globular clusters. The
sample we selected is much smaller, comprising only $48$ globular
clusters. Moreover, as explained below, data inhomogeneity has
forced us to consider separately also two smaller subsamples for which
homogeneous measurements of distance moduli are available.

Since we are aiming at a model-independent approach, we have
carefully excluded from our sources those studies that provide
exclusively model-dependent quantities (in particular, surface
brightness profiles from King model fits). Since we wish to
determine error bars on the FP coefficients, we have restricted
our attention to data with objective estimates of uncertainties.
Finally, we have given priority to homogeneous data-sets (i.e.
data from the same authors, taken with the same instruments, and
reduced in a similar fashion).

In conclusion, our sample has been selected as the intersection of
three data-sets: one with well-determined distance moduli
obtained by merging \cite{1999AJ....118.1738F} and \cite{2005A&A...432..851R},
one with well-determined line-of-sight velocity dispersions \citep{1993ASPC...50..357P}, and one with surface brightness profiles
\citep{1995AJ....109..218T} from which we could recompute apparent
magnitudes and effective radii. The properties of our sample of $48$ GCs are summarized
in Table \ref{samplestats}. In the Appendix we describe the behaviour of two smaller
sub-samples, that are obtained by considering either \cite{1999AJ....118.1738F} or
\cite{2005A&A...432..851R}, separately.

\addtocounter{table}{1}

\section{The ``Fundamental Plane"}
\label{sec:FP}

After extracting distance moduli $|m-M|_0$ and reddening values
$E(B - V)$ from our sources and reconstructing the apparent
effective radii $r_e$ and apparent integrated magnitudes $m$, we
can proceed to choose the two photometric variables that enter the
FP from the three definitions (unit-dependent constants, which play a role only in determining the zero point, are set by measuring $R_e$ in kpc, $r_e$ in arcsec, and $\mathit{SB}_e = 41.40 - 2.5 \log(L/{2\pi R^2_e})$ in mag/arcsec$^2$ where $L$ is expressed in units of the solar V-band luminosity\footnote{Our source, \cite{1995AJ....109..218T}, gives surface brightnesses in the V-band, although it is actually based on data taken in different passabands.}):

\begin{equation}
  \label{eq:1}
\log{R_e} = \frac{1}{5} {\left| m - M \right|}_0 + \log{r_e} - 7.31~,
\end{equation}

\begin{equation}
  \label{eq:2}
\mathit{SB}_e = m - 3.1 E(B - V) + 5 \log{r_e} + 1.99~,
\end{equation}

\begin{equation}
  \label{eq:3}
M = m - {\left| m - M \right|}_0 - 3.1 E(B - V)~.
\end{equation}

At this stage, some comments on the possible correlations of
errors are in order. We can be quite confident in assuming that
errors on distance moduli and reddening values do not correlate
with errors on apparent effective radii and integrated apparent magnitudes.
On the other hand, as noted earlier, some degree of
anti-correlation is present between apparent magnitudes and
effective radii \citep[see ][ for a discussion of correlations among analogous parameters of the FP of early type galaxies]{1999MNRAS.308.1037T}. From Eq.~(\ref{eq:2}) above, we see that $\mathit{SB}_e$
is the only photometric coordinate of the parameter space that
does not depend on distance moduli (although it depends on the
reddening). Therefore, as long as errors on distance moduli
dominate the error budget, for optimally dealing with the problem
of error correlations, it is appropriate to choose from the three
photometric parameters listed above any of the other two
photometric quantities together with $\mathit{SB}_e$. Even if the
contribution of the error on $\log{r_e}$ to the error on $\mathit{SB}_e$ is
comparable to that of the error on distance moduli, it is partly
cancelled by the anti-correlated error on the apparent magnitude
$m$.

These arguments confirm that the standard use of the two
photometric variables $\log{R_e}$ and $\mathit{SB}_e$ is justified and
sensible.

For an ordinary least square linear regression in the three-dimensional parameter space
defined by $\log{R_e}$, $\mathit{SB}_e$, and $\log{\sigma}$, we are now
left with choosing which coordinate to take as dependent variable,
the other two being regarded as independent variables. In the
literature the most common choice as dependent variable is $\mathit{SB}_e$, which is known to
provide ``a better and more stable fit" with respect to other
options \citep{1995ApJ...438L..29D}. In fact, different choices
give different results, and it is not clear which one is the best.
This is related to the rather narrow distribution of the cluster data
points on the FP viewed face on (i.e. to the datapoints lying more
around a line than a plane, see discussion in the following sections),
a point that we have tested by a set of simple simulations.
If the data lie approximately on a line in parameter space, then any plane passing through that line is a good fit to the data. The best fit is different when different variables are chosen as dependent variables because the sum of squared residuals along that coordinate gets minimized; in this way, the fitting algorithm chooses the best-fit plane almost at random among the infinitely many planes that pass through the line, by fitting the scatter around such line.

We find the following fit by the Fundamental Plane based on $\mathit{SB}_e$:
\begin{equation}
  \label{eq:4}
\mathit{SB}_e = (3.28 \pm 0.57) \log{R_e} - (3.59 \pm 0.39) \log{\sigma} +
(27.95 \pm 2.94)~,
\end{equation}

\noindent In turn, fitting through $\log{R_e}$ we get:

\begin{equation}
  \label{eq:5}
\log{R_e} = (0.13 \pm 0.02) \mathit{SB}_e + (0.30 \pm 0.13) \log{\sigma} -
(4.94 \pm 0.50)~,
\end{equation}

\noindent and, finally, through $\log{\sigma}$:

\begin{equation}
  \label{eq:6}
\log{\sigma} = - (0.19 \pm 0.01)\mathit{SB}_e + (0.42 \pm 0.12) \log{R_e} +
(5.07 \pm 0.46)~.
\end{equation}

In all these fits, we weighted the data-points by the inverse
squared uncertainty on the relevant dependent variable, neglecting
other uncertainties. The constant term assumes that angular
effective radii are measured in arcseconds, velocity dispersions in
kilometers per second, distance moduli and integrated apparent
luminosities are measured in magnitudes, and surface brightness is
expressed in magnitudes per square arcsecond (see also the
beginning of this Section).

The relatively large uncertainties on the fit coefficients are not
the result of overestimating error bars on data points. In fact,
the reduced $\chi^2$ for the FP fits is rather high: $9.56$,
$4.38$ and $18.88$, respectively.
These values derive either from {\it underestimating} observational
errors (which is probably not the case, given the conservative
approach taken in this paper) or, more likely, from the presence
of intrinsic scatter in the FP relationship. We will address this
issue further in the next Section, by discussing the correlations
between the residuals to the FP and other GC observational parameters.

\begin{figure}
 \resizebox{\hsize}{!}{\includegraphics[angle = 270]{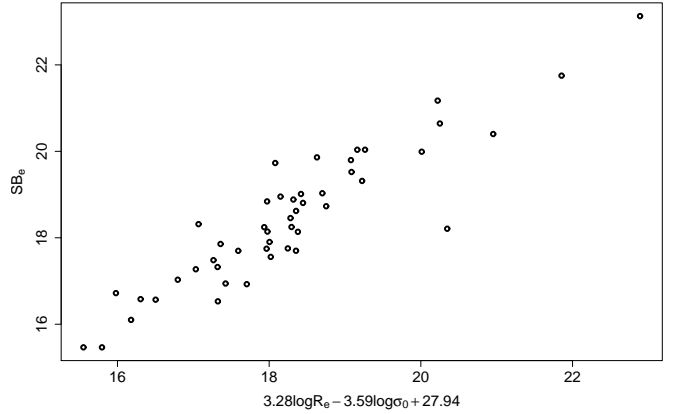}}
 \caption{Edge-on view of the FP for the galactic GCs.}
 \label{Edgeonplane}
\end{figure}

It is customary \citep[see][]{1995ApJ...438L..29D} to re-express
the FP in terms of $\log{R_e}$, even if we prefer, as argued
above, to carry out the fit through $\mathit{SB}_e$. This involves an
algebraic manipulation of the coefficients obtained from the
$\mathit{SB}_e$ based fit. Therefore, to get error bars for the derived
coefficients it is natural to apply standard error propagation
techniques, thus obtaining from Eq.~(\ref{eq:4}):

\begin{equation}
  \label{eq:7}
\log{R_e} = (1.09 \pm 0.31)\log{\sigma} + (0.30 \pm 0.05) \mathit{SB}_e -
(8.52 \pm 2.35)~,
\end{equation}

\noindent which can be easily compared to the results of other
studies or to theoretical predictions. We should recall that error
bars computed through standard error propagation techniques are
not necessarily accurate when uncertainties are large, because
such techniques are linear. Figure \ref{Edgeonplane} shows an
edge-on view of the FP. The scatter in $\log R_e$ is approximately $0.15$.

Equation (\ref{eq:4}) is easily compared to previous results, e.g. Eq. 6 in \cite{1995ApJ...438L..29D}, which we report here in our notation:
\begin{equation}
  \label{eq:DJ6}
\mathit{SB}_e = (2.9 \pm 0.1) \log{R_e} - (4.1 \pm 0.2) \log{\sigma} +
(19.8 \pm 0.1)~.
\end{equation}
The coefficients we obtain by fitting through $\mathit{SB}_e$ are compatible,
within two sigma, with the results by \cite{1995ApJ...438L..29D}, while the zero point is not, due to different conventions in measurement units.
The results recorded in Eqs. (\ref{eq:4})-(\ref{eq:7}) are not to be seen as interesting because they are consistent with those of other investigations, but rather because they demonstrate the difficulty in fitting by a plane, thus leading to the discussion of Sect. \ref{sect:3.1}

We finally note that the FP coefficients just found in Eq. (7) by fitting through $\mathit{SB}_e$ and inverting for $\log{R_e}$ are consistent
with those of the FP for early-type galaxies, also in the zero-point
constant, namely \citep{1996MNRAS.280..167J}:

\begin{equation}
  \label{eq:8}
\log{R_e} = 1.24 \log{\sigma_0} + 0.33 \mathit{SB}_e - 8.895~,
\end{equation}

\noindent with reported scatter of $0.08$ in $\log R_e$, but the error bars are marginally compatible even with the zero-tilt plane:

\begin{equation}
  \label{eq:9}
\log{R_e} = 2.0 \log{\sigma} + 0.40 \mathit{SB}_e + \gamma~.
\end{equation}

This result shows that better data are required in order to draw convincing
physical conclusions.

\subsection{A Fundamental Line?}
\label{sect:3.1}

If on the FP viewed
face-on the data-points are located mainly within a thin strip (with
width comparable to the scatter around the FP) it may be difficult to discriminate between a plane
distribution and a {\it line distribution}. In principle we
could argue that, by suitable rescaling of the coordinates, any
plane distribution may be stretched into a line distribution and
vice-versa, although we should check how the coordinate
transformation changes the relevant error bars. Rather than
following this apparent paradox as a general problem of
statistics, here below we adopt the view that the standard choice
of coordinates used in discussions of the FP has its own physical
justification and wish to check whether indeed, in the standard FP
coordinates, the set of points occupies a rather slim cylindrical
region of parameter space or not. We thus calculate the
eigenvalues and eigenvectors of the tensor of inertia of the
unweighted data-points in the $(\mathit{SB}_e - \langle \mathit{SB}_e \rangle, \log{R_e} -
\langle \log{R_e} \rangle, \log{\sigma}
- \langle \log{\sigma} \rangle)$ coordinates (i.e. in the ``center
of mass" frame) and obtain for the eigenvalues $I_1$, $I_2$, and $I_3$:

\begin{equation}
  \label{eq:10}
I_1 = 1.99
\end{equation}

\begin{equation}
  \label{eq:11}
I_2 = 120.68
\end{equation}

\begin{equation}
  \label{eq:12}
I_3 = 121.33
\end{equation}

\noindent and for the respective eigenvectors (in the aforementioned coordinates):

\begin{equation}
  \label{eq:13}
v_1 = (0.98, 0.09, -0.15)
\end{equation}

\begin{equation}
  \label{eq:14}
v_2 = (0.05, 0.67, 0.74)
\end{equation}

\begin{equation}
  \label{eq:15}
v_3 =  (0.17, -0.73, 0.66)
\end{equation}

The eigenvector corresponding to the smallest eigenvalue sets the
direction of the ``Fundamental Line". This procedure is equivalent
to unweighted least square fitting in terms of a Fundamental Line,
(i.e. finding the straight line which minimizes the sum of distances
to the points squared) but gives also a measure of the degree of
{\it axisymmetry} of the distribution of data-points around such
Fundamental Line, in terms of the difference between the other two
eigenvalues $I_2$ and $I_3$:

\begin{equation}
  \label{eq:16}
R \equiv \frac{2 |I_2 - I_3|}{(I_2 + I_3)} = 0.005,
\end{equation}

\noindent {\it which is remarkably small}.

Indeed, it was already noted \citep{1997AJ....114.1365B, 1998NewA....3..219B} that the galactic GCs lie close to a straight line in the $\log{R_e}$,
$\mathit{SB}_e$, $\log{\sigma}$ parameter space. But, to the best of our
knowledge, the present paper is the first to identify and quantify the degree of axisymmetry around such line, a fact that demands an adequate theoretical explanation.
Since \cite{1998NewA....3..219B} suggests that
``at earlier times, globular clusters populated a line in the
three-dimensional S-space, i.e. their original dynamical structure
was fully determined by a single physical parameter", we also
looked for correlations between residuals to the Fundamental Line
and GC relative age \citep[taken from][]{2005AJ....130..116D},
finding only a very small, but positive, correlation coefficient
of $0.05$, which further decreases to $0.03$ if more robust
non-parametric correlation estimators (such as the Spearman
estimator \citep[e.g., see][]{LarryWasserman}) are used.
However, this apparently negative result is not conclusive, because
our sample contains mainly metal-poor clusters (see Sect.~\ref{selection}),
and \cite{2005AJ....130..116D} clearly state that ``the age
dispersion for the metal-poor clusters is 0.6 Gyr (rms),
consistent with a null age dispersion". In conclusion, to clarify
this point a larger sample would be desired.

\subsection{A pure photometric scaling law?}
We note that a Fundamental Line relation of the type identified above would imply the existence of a {\it pure photometric} scaling law approximately
of the form

\begin{equation}
 \label{eq:17}
\log R_e \propto \frac{1}{10} \mathit{SB}_e,
\end{equation}

\noindent i.e.
\begin{equation}
 \label{eq:18}
L \propto R^{-2}_e.
\end{equation}

\noindent This result is quite unexpected because the trend suggested by Eq. (\ref{eq:18}) would be qualitatively opposite to that of the \cite{1964ApJ...139..284F} law. In addition, we should recall that, on a sample of $143$ GCs, \cite{1994AJ....108.1292D} conclude that no significant correlation exists between integrated absolute magnitude and the logarithm of the effective radius. From the relevant plot in their Fig. 6, we note that an underlying trend might be masked by the presence of two or three outlier points; furthermore, in their Table 1 \cite{1994AJ....108.1292D} actually report a positive correlation (with $r = 0.819$) between $\log{R_e}$ and $\mathit{SB}_e$ (${\langle\mu_V\rangle}_h$ in their notation), which can be recast into a scaling law of the form of Eq. (\ref{eq:18}), recovering the correct sign for the exponent. 

On the other hand, the correlation that we find in the photometric plane turns out to be in qualitative agreement with the general trend noted for elliptical galaxies by \cite{1977ApJ...218..333K}; see also \cite{1987IAUS..127..379H, 1992MNRAS.259..323C}. We have checked that, for the sample of 88 GCs obtained by intersecting the two samples of \cite{1999AJ....118.1738F} and \cite{2005A&A...432..851R} with that of \cite{1995AJ....109..218T}, the correlation that we find in the photometric plane indeed has a consistent slope but a different zero point with respect to the relation $\mu_{eV} = 2.94 logR_e + 19.48$ reported by Hamabe \& Kormendy. To be sure,  the $\mu_{eV}$ variable is very sensitive to the quality of the surface brightness profiles; therefore, a direct detailed comparison with the least squares regression of \cite{1977ApJ...218..333K} in the $\mu_{eV}$ variable cannot be performed easily and would require additional discussion.

In fact, the value of the exponent in Eq. (\ref{eq:18}) is very sensitive to the adopted fitting procedure. The projected Fundamental Line is not equal to the
straight line that minimizes the sum of the squared residuals along one coordinate in the plane $(\log{R_e}, \mathit{SB}_e)$, i.e. it does not coincide with the line one would get with simple linear least squares fitting. To recover the same relation as predicted by the projected Fundamental Line,
one should perform an impartial regression in the $(\log{R_e}, \mathit{SB}_e)$ plane, i.e. one should minimize the sum of squared distances to the line, not just the residuals in one coordinate. Moreover, it should be noted that the least squares method implicitly used for fitting the Fundamental Line is non-robust, i.e. its results may be excessively influenced by outliers; this is obviously true also for the Fundamental Line projection and the derived scaling relation.

The scatter around the Fundamental Line has the potential for significantly blurring the univariate correlation obtained by projecting the Fundamental Line on the photometric $(\log{R_e}, \mathit{SB}_e)$ plane. Therefore, we anticipate that either a narrowing of the sample to a smaller set of GCs or the identification of a variable $\eta$ capable of reducing that scatter \citep[$\eta$ could be the central slope of the surface brightness profile, as in the case of the FP, or age as suggested by][]{1998NewA....3..219B} might lead to a much improved univariate relation between $\log{R_e}$ and $\mathit{SB}_e$. Possibly, a new bivariate correlation could be sought for in the $(\log{R_e}, \mathit{SB}_e, \eta)$ space. 

A clear interest in a scaling law of the form of Eq. (\ref{eq:18}) lies in the fact that its study does not require the acquisition of kinematical information. Therefore, we should be able to test it on wider samples and possibly on globular cluster systems in external galaxies. This will soon
be addressed in a follow-up paper. The exclusion of the $\log \sigma$ coordinate would also allow us to consider samples definitely more homogeneous
from the statistical point of view. Finally, the existence of a pure photometric scaling relation suggests that the physical basis for the empirical scaling should be sought more in the context of stellar populations than within the virial theorem and dynamical arguments.

\section{Two unexpected features} \label{features}

\subsection{Correlation of residuals with the central slope of the surface
brightness profile}

The scatter around the FP, as quantified by the value of the reduced $\chi^2$,
significantly greater than unity, might be interpreted as an indication of the role of nonlinearities or of a fourth parameter.

First of all, we have looked for residual correlations with the coordinates defining the space in which the FP lives, namely $\log{R_e}$, ${\mathit{SB}}_e$, and $\log{\sigma}$. By looking at the relevant
scatter plot, the residuals show no recognizable pattern with either $\log{R_e}$, or ${\mathit{SB}}_e$, or $\log{\sigma}$; from a qualitative point of view, this suggests that nonlinear corrections to the FP, if present, are unobservable, at least with present-day data quality. Quantitatively, the null hypothesis that the coefficients of nonlinear terms in the form ${(\log{R_e})}^2$, $\log{R_e}\log{\sigma}$, \ldots, be equal to $0$ cannot be refuted even at a $20\%$ significance level.

Then we have moved on to consider correlations with other
quantities, mainly of dynamical interest. In particular, we have
referred to: King model concentration parameter, central
slope of the surface brightness profiles \citep[as defined and measured
by ][]{2006AJ....132..447N}, position in the Galaxy (i.e.
galactocentric distance and vertical distance from the galactic
disk), age, relaxation time referred to the effective radius,
metallicity $[Fe/H]$, and colour $B - V$. In this respect, our
investigation is not complete, because other quantities, in
particular colour-magnitude-diagram related quantities, might have
been addressed.

\begin{figure}
 \resizebox{\hsize}{!}{\includegraphics[angle = 270]{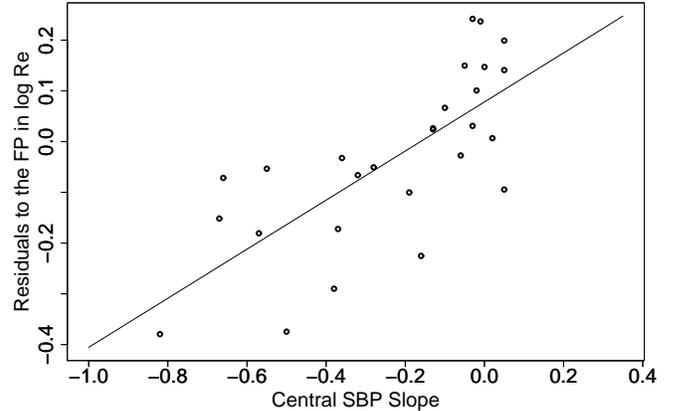}}
 \caption{Correlation of the FP residuals with
 slope of the central surface brightness profile as defined
 and measured by \protect\cite{2006AJ....132..447N}.}
 \label{Correlation}
\end{figure}

Consistent with indications from previous studies \citep[e.g., see][]{1994AJ....108.1292D}
which excluded correlations between GC properties, even in the case of
FP residuals no significant correlations
are found, with just one interesting exception, that is a statistically
significant correlation of the FP residuals with the central slope of
the surface brightness profile. In view of the results obtained by \cite{2000ApJ...539..618M} \citep[see also ][]{2007AJ....133.2764B}, a trend with galactocentric position would be expected; the fact that the present analysis is unable to detect it could be due to the selection effect in galactocentric distance characterizing our sample. Figure \ref{Correlation}
shows a plot of FP residuals against the central slopes taken from
\cite{2006AJ....132..447N}.
The trend of smaller effective radii with increasing slopes is
evident from the scatter plot; the value of the related
correlation coefficient is $0.73$. The presence of such a
correlation reinforces the view that the origin of the FP scatter
is indeed intrinsic. This result is particularly interesting, because it relates
central to global properties of these stellar systems, much like
the $M_{BH}$-$\sigma$ relationship \citep{2000ApJ...539L...9F,
2000ApJ...539L..13G} does for bulges and elliptical
galaxies. Indeed, an analogous argument has been proposed by \cite{1997AJ....114.1771F} in their study of
the core regions of elliptical galaxies and spiral bulges; in particular,
\cite{1997AJ....114.1771F} state that "Cores follow a fundamental plane
that parallels the global fundamental plane for hot galaxies but is 30\%
thicker. Some of this extra thickness may be due to the effect of massive
black holes (BHs) on central velocity dispersions". We note that in
some GCs the gravitational sphere of influence of an intermediate mass
black hole (for example, such a black hole has been argued by \cite{2008arXiv0801.2782N} to be present in Omega Cen) could have an
angular size comparable to the slit size used for integrated light
measurements of the velocity dispersion. This effect might be
at the origin of the observed correlation; fitting cuspy GC cores with double-power law profiles, \citep[see][]{1996AJ....111.1889B} or with simple dynamical models \citep{1994A&A...288...43C} might shed further light on this question.

We have also checked whether the scatter in the FP can be related
to the presence, in our sample, of objects that are suspected to
be non-genuine galactic GCs, such as members of GC streams
\citep{1995MNRAS.275..429L}, which are probable remnants of GC
systems tidally stripped from a smaller galaxy accreted by the
Milky Way. We have compared the distribution of residuals to the
FP for those GCs that are part of such streams \citep[according to ][]{2007ChJAA...7..111G} with that of the rest of our sample
through a Kolmogorov-Smirnov test and found no evidence of significant
differences. Therefore, we argue that GC streams are not
responsible for the excessive FP scatter, or, in other words, that,
in this context, stream GCs are dynamically similar to native GCs.

\subsection{Distribution of effective virial coefficients}
\label{Sect:distkv}

\begin{figure}
 \resizebox{\hsize}{!}{\includegraphics[angle = 270]{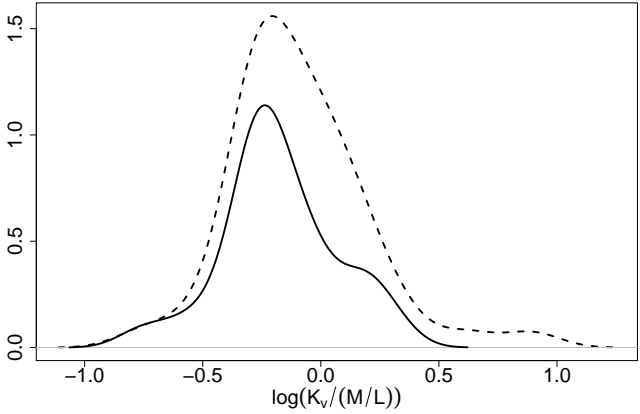}}
 \caption{Probability density distribution of $\log{K_V/(M/L)}$ for clusters with velocity dispersions measured in integrated light (solid line) and for the entire sample (dashed line), reconstructed by application of the non-parametric method of kernel density estimation. Units have been chosen in such a way that $K_V/(M/L) = 1$ for NGC 104.}
 \label{KDEkvs}
\end{figure}

We have reconstructed the distribution of $K_V/(M/L) \equiv {G L}/{(R_e \sigma^2)}$ \citep[virial
coefficient divided by the mass-to-light ratio; for a thorough discussion, see][]{2002A&A...386..149B} for our sample,
making use of kernel density estimation techniques \citep[see][]{LarryWasserman}. In
many respects, these non-parametric techniques are an extension of
histograms, best suited to spot structure that histograms may
hide. The $K_V/(M/L)$ distribution looks bimodal, even if we choose a slightly
larger bandwidth than optimal ($1.2$ times) so to err on the side of oversmoothing
(undersmoothing could easily produce fictitious bi- or multi-modal distributions).
The indication of bimodality is stronger if the analysis is restricted
to the $28$ systems for which the line-of-sight velocity
dispersion is obtained from integrated light measurements, while for the entire sample of $48$ GCs it almost disappears. Figure
\ref{KDEkvs} illustrates the reconstructed probability density
distribution of $\log{K_V/(M/L)}$ for the restricted and the unrestricted case, with the restricted case exihibiting
the superposition of a narrow peak around a negative value onto a
broader distribution with positive mean. Therefore we note that inhomogeneity of velocity dispersion measurements is not responsible for the bimodality, which appears more clearly when the sample is restricted to integrated light velocity dispersion measurements only. This result is also suggestive of an interpretation in terms of the presence of intermediate mass
black holes, which would naturally affect the value of the central velocity dispersion (best diagnosed by the integrated light measurements), in some of the globular clusters, but, as for the feature noted in the previous subsection, a thorough investigation on a larger sample would be required in order to make a fully convincing case.
Even though the mass-to-light ratio of a stellar population may be heavily influenced by its metallicity, we feel confident in excluding a metallicity-dependent effect as responsible for the bimodality of the $\log{K_V/(M/L)}$ because no correlation is observed between effective virial coefficient and metallicity; using $[Fe/H]$ data from the McMaster Catalogue, we find a correlation coefficient of $0.04$ between these two quantities, which is statistically compatible with zero.

In addition, this result makes a point which is clearly against
the zero tilt FP scenario, by showing that the $\log{K_V/(M/L)}$ values have
a non-trivial distribution, i.e. inconsistent with a sharply peaked distribution broadened into a bell-shaped curve by observational errors.

\section{Selection effects} \label{selection}
 
As noted earlier, the McMaster Catalogue of Milky Way Globular Cluster Parameters
\citep{1996AJ....112.1487H} lists $151$ globular clusters, while the
sample considered in the present paper is much smaller, comprising $48$ GCs. By comparison, the sample used by
\cite{1995ApJ...438L..29D} contains $55$ globular clusters, those
for which a measurement of line-of-sight velocity dispersions is
available in \cite{1993ASPC...50..357P}. Under these conditions a
thorough study of selection effects is mandatory.
 
We proceeded as follows. First of all we defined a list of
relevant parameters, extracted from those tabulated in the 2003
version of the McMaster Catalogue, namely the distance of the GC
from the Galactic Center $R_{gc}$, the vertical distance to the
Galactic Plane $Z$, the distance from the Sun $d_{\odot}$, the
absolute integrated magnitude $M_V$, the integrated colour $B-V$
(uncorrected for reddening), the metallicity as measured by the
$[Fe/H]$ parameter, the King model concentration parameter $c$,
the projected effective radius $R_e$, the logarithm of
the relaxation time referred to the effective radius $\log{t_e}$,
and the central surface brightness $\mu_V$. We then compared the
distribution of these parameters for our sample with that of the
complementary sample, that is of the sample containing all of the
GCs in the McMaster catalogue except those of our sample.
 
In assessing the importance of selection effects, we used the
quantities taken from the McMaster Catalogue; we did not try to
recompute any of them anew for consistency reasons, because we
would otherwise treat our sample (for which quality data are
available) in a different way with respect to the rest of the
clusters.
 
The comparison has been performed through the non-parametric
Kolmogorov-Smirnov test. The test gives an estimate of
the probability of the sample and of its complementary set being
drawn from the same distribution. If this probability is extremely
low, say less than ${10}^{-3}$, we can quite confidently assert
that a selection effect is present, i.e. that the sample we
considered is not statistically representative of the whole
population of galactic GCs. Being non-parametric, this test does
not assume a given functional dependence for the distribution
function of the data, which is unknown in principle; this is an
important requirement in this context, because the distribution
functions of the various parameters we considered are often far
from gaussian and may be highly skewed or have long tails.
 
We performed a similar test also for the wider sample of $55$ GCs
with velocity dispersions measured by \cite{1993ASPC...50..357P}. \cite{1995ApJ...438L..29D} assumes no relevant selection effects on
that sample, at variance with our analysis, that shows that
some selection effects are indeed present at a significant level.

\begin{table}
\label{oursample}
 \caption{Comparison of the $48$ GC
sample of this paper with the complementary sample within the
McMaster Catalogue, from which the data listed in this Table are
taken. Kolmogorov-Smirnov p-values are approximated to one
significant digit, while all other numbers are truncated to the
second digit past the decimal dot.}
\begin{center}
\begin{tabular}{llllll}
\hline\hline
ID & ${p_{K\&S}}^a$ & ${\mu_s}^b$ & ${\mu_c}^c$ & ${\sigma_s}^d$ & ${\sigma_c}^e$\\
\hline
$R_{gc}$ (kpc)& $0.2$ & $10.05$ & $12.96$ & $13.62$ & $22.02$\\
$Z$ (kpc)& $0.1$ & $1.94$ & $0.53$ & $8.13$ & $18.09$\\
$d_\odot$ (kpc)& $0.2$ & $11.72$ & $16.80$ & $12.41$ & $20.60$ \\
$M_V$ & $0.0002$ & $-7.78$ & $-6.69$ & $1.07$ & $1.58$\\
$B-V$ & $0.004$ & $0.90$ & $1.18$ & $0.27$ & $0.48$\\
$[Fe/H]$ & $0.1$ & $-1.45$ & $-1.22$ & $0.49$ & $0.58$\\
$c$ & $0.0006$ & $1.82$ & $1.44$ & $0.54$ & $0.54$\\
$R_e$ (pc)& $0.04$ & $3.63$ & $4.94$ & $3.22$ & $4.68$\\
$\log{t_e}$ (yr)& $0.2$ & $9.09$ & $9.04$ & $0.39$ & $0.54$\\
$\mu_V$ & $8 \cdot 10^{-7}$ & $16.99$ & $19.82$ & $2.01$ & $2.66$\\
\hline\hline
\end{tabular}\\
\end{center}
$^a$ p-value for the null hypothesis of the sample and its complementary sample coming from the same distribution\\
$^b$ Mean value of the relevant parameter on the sample\\
$^c$ Mean value of the relevant parameter on the complementary sample\\
$^d$ RMS of the relevant parameter on the sample\\
$^e$ RMS of the relevant parameter on the complementary sample\\
\end{table}
\begin{table}
\caption{Comparison of the $55$ GC
sample of Djorgovski (1995) with the complementary sample within the
McMaster Catalogue, from which the data listed in this Table are
taken. Kolmogorov-Smirnov p-values are approximated to one
significant digit, while all other numbers are truncated to the
second digit past the decimal dot. See Table 2 for column heading explanations.}
\label{theirsample}
\begin{center}
\begin{tabular}{llllll}
\hline\hline
ID & ${p_{K\&S}}$ & $\mu_s$ & $\mu_c$ & $\sigma_s$ & $\sigma_c$\\
\hline
$R_{gc}$ (kpc)& $0.1$   & $9.70$    & $13.38$ & $12.90$ & $22.72$ \\
$Z$      (kpc)   & $0.5$   & $1.54$    & $0.66$  &$7.71$  & $18.73$ \\
$d_\odot$ (kpc)  & $0.1$   & $11.33$   & $17.39$ & $11.94$ & $21.13$ \\
$M_V$       & $2 \cdot {10}^{-5}$ & $-7.85$ & $-6.57$& $1.15$ & $1.52 $\\
$B-V$       & $0.002$ & $0.90$    & $1.21$ & $0.26$  & $0.49 $\\
$[Fe/H]$    & $0.02$  & $-1.48$   & $-1.19$& $0.46$  & $0.59 $\\
$c$         & $0.0009$& $1.79$    & $1.43$ & $0.52$  & $0.56 $\\
$R_e$  (pc) & $0.04$  & $3.62$   & $5.05$& $3.04$ & $4.84$\\
$\log{t_e}$ (yr)& $0.2$   & $9.11$    & $9.02$ & $0.40$  & $0.54 $\\
$\mu_V$     & $2 \cdot {10}^{-8}$ & $16.96$ & $20.06$ & $1.94$& $2.58 $\\
\hline\hline
\end{tabular}\\
\end{center}
\end{table}
Our statistical tests are summarized in Tab. 2 (our
sample against the rest of galactic GCs) and
Tab.~\ref{theirsample} (the \cite{1993ASPC...50..357P} sample
against the rest of galactic GCs), which also list the mean and
standard deviation of the chosen sample and of its complementary sample.
 
The results of our statistical analysis can be summarized as
follows. Selection effects are definitely present, although they
do not necessarily hinder the conclusions of the present paper.
Both samples (\cite{1995ApJ...438L..29D} and ours) are biased in
favor of clusters relatively close to the Sun; both are biased
towards intrinsically more luminous and centrally brighter clusters. These
selection effects, although expected, must be kept in mind when
addressing issues such as whether the FP properties change with
Galactocentric distance.
 
On the other hand, it is somewhat surprising to find that both
samples are biased in the direction of bluer and more centrally
concentrated clusters (even though this latter feature may partly
result from the fact that the McMaster Catalogue assigns a concentration parameter of $2.5$ to GCs recognized to
be in the Post Core Collapse phase of their evolution). To be sure,
we are aware that concentration parameters show a trend with GC
integrated luminosities \citep[see]{1994AJ....108.1292D},
so the bias towards concentrated clusters might be a byproduct
of the selection effect in magnitude.
Our sample has somewhat lower mean metallicity than average;
this selection effect becomes even stronger if only the subsample
with distance moduli from \cite{1999AJ....118.1738F} is used (see Appendix).
 
The clusters in the \cite{1995ApJ...438L..29D} sample are slightly
smaller than average, in terms of $R_e$, while for both samples
there seems to be no bias in the relaxation time referred to
$R_e$. Therefore, we feel confident in excluding major dynamical
differences between the two samples (\cite{1995ApJ...438L..29D}
and ours) and the overall population of galactic GCs, which makes
it reasonable to extend our conclusions to the whole system of
galactic GCs.
 
In closing this Section we should mention that selection effects
might be present in more than one variable taken together, even
when they do not show up by considering only one variable at a
time, as we did. That is, if $x$ and $y$ are two variables, the
distribution function of our sample could be different from that
of the rest of the GCs on the $(x,y)$ plane, while being
indistinguishable when projected onto the $x$ or the $y$ axes.
This issue has been considered to be beyond the scope of the
present paper.

\section{Summary, discussion, and conclusions}
 
\subsection{Results}
\label{risultati}
 
Based on the identification of an optimal sample of 48 globular
clusters selected in terms of homogeneity criteria for the
photometric data available in the literature, the most important
results of the present study are the following:
 
\begin{itemize}
 
\item We have determined the coefficients of the FP for the
galactic globular cluster system, {\it with error bars} (see
Eq.~\ref{eq:7}).
 
\item Since the reduced $\chi^2$ of the fit has been found to be
significantly greater than unity, we conclude that either we have
underestimated errors or, more likely, that the relatively large
scatter around the FP (of about $0.15$ in $\log{R_e}$) is intrinsic. We
found that the so-called GC streams are likely to be not
responsible for the scatter around the FP.
 
\item We have shown that in the standard FP coordinates
the set of points representing our sample of GCs occupies a rather
slim cylindrical region of parameter space, which suggests that
the relevant scaling relation might be around a line, rather than a plane;
this confirms a result noted previously by Bellazzini (1998). We
think that this is the origin of the difficulties in fits by a
plane noted earlier \citep[e.g., see][]{1995ApJ...438L..29D, DeMicheliTesi}. In addition, we have derived eigenvalues and
eigenvectors for the tensor of inertia of the distribution of
data-points in the $\mathit{SB}_e$, $\log{R_e}$, $\log{\sigma}$
coordinates and found that such distribution is remarkably
axisymmetric around a line, mainly along $\mathit{SB}_e$.
 
\item For our sample, we found no statistically significant
correlation of FP residuals with metallicity, colour,
concentration parameter, position in the Galaxy, ages, and
relaxation times, but we found a potentially interesting
correlation with the central slope of the surface brightness
profile.
 
\item By means of kernel density estimation techniques, we have
reconstructed the probability density distribution of the
effective virial coefficient $K_V/(M/L)$ of GCs, providing
evidence for bimodality. We argue that this feature is not related
to the standard disk/halo dichotomy, because our sample comprises
only clusters with metallicity below about $-0.8$, which are
therefore likely to be part of the halo population.
 
\item By means of non-parametric statistical tests, we studied the
selection effects affecting our sample and the sample used by
\cite{1995ApJ...438L..29D} in his pioneering paper on the FP for
the galactic GCs, demonstrating that some selection effects are
present (in particular, our sample is biased in the direction of
bright, metal poor clusters close to the Sun). No significant selection
effects are present in the relaxation time referred to the
effective radius, which is reassuring because it suggests that our
sample is not dynamically different from the remaining population
of galactic GCs.
 
\end{itemize}
 
\subsection{Discussion and conclusions}
 
The error bars on the FP coefficients are rather large. Therefore,
on the basis of currently available data it is difficult to tell
whether indeed the FP relation of GCs is a continuation of that of
early-type galaxies. The limited luminosity range of galactic GCs
may be an issue here; studies of extragalactic GC systems such as that
of NGC 5128, which reach a higher upper luminosity limit \citep[e.g., see ][]{2007A&A...469..147R, 2004ApJ...610..233M}, are in line with the present study.
In principle, the current FP fit does not exclude (at least
at a marginal level) that a no-tilt relation is
at the basis of the observed correlations. On the other hand, we
found concrete indications that the shape of the distribution of
$K_V/(M/L)$ is bimodal; if such indications will be confirmed,
this would clearly point against the naive interpretation that the
FP just reflects the virial constraint.
 
The hints of a bimodal distribution of the effective virial
coefficient $K_V/(M/L)$ brought us to consider one specific
physical ingredient that might play an important role in the
statistical properties of the GC system, that is the possible
presence of intermediate mass black holes. In fact, such presence
would also have a counterpart in the second unexpected feature
noted in our statistical investigation, i.e., the trend of FP
residuals with the central slope of surface brightness profiles.

As to the possibility that the Fundamental Plane should be replaced,
for globular clusters, by a Fundamental Line relation, before trying
to put forward a physical interpretation we prefer to wait for
confirmations from a wider sample or from studies of globular cluster
systems in external galaxies. The possibility of getting soon an answer
to this issue is reasonable because, as discussed at the end of Sect. 3,
the Fundamental Line relation appears to imply the existence of a pure
photometric scaling law.

\emph{Acknowledgements.} We wish to thank L. Ciotti, S. Degl'Innocenti, F. Ferraro, M. Lombardi, E. Noyola, G. Punzi, S. Trager, M. Trenti, and E. Vesperini for their comments and helpful suggestions. We also thank the Referee, for making a number of constructive remarks that have helped us improve the paper.
The initial part of the work started in order to clarify some questions raised in the Thesis by \cite{DeMicheliTesi}.
This research has made use of the SIMBAD data-base, operated at CDS, Strasbourg, France.
This work was partially supported by the Italian MiUR.

\appendix
\section{Discussion of the FP on two subsamples with uniform distance moduli}

For the sub-sample of $35$ GCs with distance moduli from \cite{1999AJ....118.1738F} we find
the following fit by the Fundamental Plane based on $\mathit{SB}_e$:
 
 \begin{equation}
   \label{eq:104}
 \mathit{SB}_e = (3.2 \pm 0.7) \log{R_e} - (3.4 \pm 0.5) \log{\sigma} +
 (27.96 \pm 3.5)~.
 \end{equation}
 
 \noindent In turn, fitting through $\log{R_e}$ we get
 
 \begin{equation}
   \label{eq:105}
 \log{R_e} = (0.14 \pm 0.03) \mathit{SB}_e + (0.3 \pm 0.1) \log{\sigma} -
 (5.04 \pm 0.6)~,
 \end{equation}
 
 \noindent and, finally, through $\log{\sigma}$:
 
 \begin{equation}
   \label{eq:106}
 \log{\sigma} = - (0.18 \pm 0.02)\mathit{SB}_e + (0.4 \pm 0.2) \log{R_e} +
 (5.15 \pm 0.6)~.
 \end{equation}
 
 \noindent In turn, for the sub-sample of $34$ GCs with distance moduli from  \cite{2005A&A...432..851R} we find
 the following fit by the Fundamental Plane based on $\mathit{SB}_e$:
 
 \begin{equation}
   \label{eq:107}
 \mathit{SB}_e = (2.66 \pm 0.75) \log{R_e} - (4.02 \pm 0.47) \log{\sigma} +
 (31.70 \pm 3.9)~.
 \end{equation}
 
 \noindent In turn, fitting through $\log{R_e}$ we get
 
 \begin{equation}
   \label{eq:108}
 \log{R_e} = (0.12 \pm 0.03) \mathit{SB}_e + (0.26 \pm 0.18) \log{\sigma} -
 (5.19 \pm 0.65)~,
 \end{equation}
 
 \noindent and, finally, through $\log{\sigma}$:
 
 \begin{equation}
   \label{eq:109}
 \log{\sigma} = - (0.17 \pm 0.02)\mathit{SB}_e + (0.36 \pm 0.14) \log{R_e} +
 (5.16 \pm 0.52)~.
 \end{equation}

 \noindent These results are compatible with those obtained using the global sample of $48$ GCs considered in the main text of the paper. Since \cite{2005A&A...432..851R} and \cite{1999AJ....118.1738F} use a different definition of the ZAHB level, these results show that our conclusions (and in particular the FP coefficients) are relatively independent of the details of ZAHB fitting.
 
 We have also checked that the general results listed in Sect. \ref{risultati} hold for the two subsamples if studied separately.

\bibliographystyle{aa}
\bibliography{manuscript}

\longtab{1}{
\begin{longtable}{lllllllll}
\caption{\label{samplestats} Properties of the sample considered in this paper.}\\
\hline\hline
ID & ${\sigma}^a$ & $m^b$ & ${\log{r_e}}^c$ & ${|M - m|}^d_0$ & $\delta {|M - m|}^e_0$ & ${E(B-V)}^f$ & ${{d\log{\Sigma}}/{d\log{r}}}^g$ & Distance sample$^h$\\
\hline
\endfirsthead
\caption{continued.}\\
\hline\hline
ID & ${\sigma}^a$ & $m^b$ & ${\log{r_e}}^c$ & ${|M - m|}^d_0$ & $\delta {|M - m|}^e_0$ & ${E(B-V)}^f$& ${{d\log{\Sigma}}/{d\log{r}}}^g$ & Distance sample$^h$\\
\hline
\endhead
\hline\hline
\endfoot
NGC 104 & $9.72$ & $4.00$ & $2.22$ & $13.27$ & $0.08$ & $0.04$&$0.00\pm0.04$ &R05 F99\\          
NGC 288 & $2.61$ & $8.45$ & $2.16$ & $14.67$ & $0.30$ & $0.03$& & F99\\             
NGC 362 & $6.15$ & $6.26$ & $1.69$ & $14.69$ & $0.10$ & $0.05$& &R05 F99\\          
NGC 1851 & $10.21$ & $7.32$ & $1.46$ & $15.44$ & $0.09$ & $0.02$&$-0.38\pm0.11$ &R05 F99\\         
NGC 1904 & $4.59$ & $7.70$ & $1.93$ & $15.67$ & $0.08$ & $0.01$&$-0.03\pm0.07$ &R05 F99\\          
NGC 2419 & $2.66$ & $10.46$ & $1.17$ & $19.88$ & $0.29$ & $0.03$& & F99\\           
NGC 2808 & $14.04$ & $6.42$ & $1.77$ & $14.96$ & $0.13$ & $0.23$&$-0.06\pm0.07$ &R05 F99\\         
NGC 3201 & $4.30$ & $7.21$ & $2.30$ & $13.58$ & $0.13$ & $0.21$& &R05 F99\\         
NGC 4147 & $2.54$ & $10.32$ & $1.46$ & $16.42$ & $0.10$ & $0.02$& &R05 F99\\        
NGC 4590 & $2.39$ & $8.32$ & $1.96$ & $15.16$ & $0.07$ & $0.04$& &R05 F99\\         
NGC 5053 & $1.15$ & $9.79$ & $2.29$ & $16.17$ & $0.19$ & $0.03$& & F99\\            
NGC 5272 & $5.22$ & $6.65$ & $1.83$ & $14.99$ & $0.14$ & $0.01$&$-0.05\pm0.10$ & F99\\             
NGC 5286 & $8.00$ & $7.41$ & $1.90$ & $15.25$ & $0.29$ & $0.24$&$-0.28\pm0.11$ & F99\\             
NGC 5466 & $1.57$ & $9.11$ & $2.13$ & $16.12$ & $0.29$ & $0.00$& & F99\\            
NGC 5694 & $5.50$ & $10.17$ & $1.43$ & $17.80$ & $0.11$ & $0.09$&$-0.19\pm0.11$ &R05 F99\\         
NGC 5824 & $10.73$ & $8.90$ & $1.44$ & $17.63$ & $0.11$ & $0.13$&$-0.36\pm0.16$ &R05 F99\\         
NGC 5904 & $5.24$ & $5.93$ & $1.97$ & $14.44$ & $0.09$ & $0.03$&$0.05\pm0.07$ &R05 F99\\          
NGC 5946 & $3.70$ & $9.74$ & $1.62$ & $15.41$ & $0.29$ & $0.55$& &R05 \\            
NGC 6093 & $13.32$ & $7.47$ & $1.56$ & $15.09$ & $0.15$ & $0.18$&$-0.16\pm0.07$&R05 F99\\         
NGC 6121 & $3.59$ & $5.67$ & $2.34$ & $11.62$ & $0.30$ & $0.36$& & F99\\            
NGC 6171 & $3.67$ & $8.29$ & $2.16$ & $14.07$ & $0.17$ & $0.33$& &R05 F99\\         
NGC 6205 & $6.62$ & $6.00$ & $1.96$ & $14.45$ & $0.08$ & $0.02$&$-0.10\pm0.15$&R05 F99\\          
NGC 6218 & $3.92$ & $7.27$ & $2.02$ & $13.66$ & $0.23$ & $0.18$& &R05 F99\\         
NGC 6254 & $5.48$ & $6.70$ & $2.08$ & $13.35$ & $0.29$ & $0.28$&$0.05\pm0.07$& F99\\             
NGC 6256 & $6.50$ & $10.91$ & $1.89$ & $14.97$ & $0.37$ & $0.84$& &R05 \\           
NGC 6266 & $14.23$ & $6.50$ & $1.81$ & $14.31$ & $0.21$ & $0.47$&$-0.13\pm0.08$&R05 F99\\         
NGC 6284 & $6.20$ & $8.82$ & $1.70$ & $16.08$ & $0.21$ & $0.28$&$-0.55\pm0.14$&R05 \\             
NGC 6293 & $7.60$ & $8.28$ & $1.76$ & $14.88$ & $0.24$ & $0.39$&$-0.67\pm0.08$& R05\\             
NGC 6325 & $5.80$ & $10.30$ & $1.88$ & $14.82$ & $0.39$ & $0.89$& & R05\\           
NGC 6341 & $5.00$ & $6.53$ & $1.69$ & $14.74$ & $0.29$ & $0.02$&$-0.01\pm0.04$& F99\\             
NGC 6342 & $4.60$ & $9.80$ & $1.68$ & $15.00$ & $0.26$ & $0.44$& &R05 \\            
NGC 6366 & $0.93$ & $9.27$ & $2.26$ & $12.88$ & $0.30$ & $0.69$& & F99\\            
NGC 6362 & $2.76$ & $7.71$ & $2.24$ & $14.45$ & $0.16$ & $0.09$& &R05 \\            
NGC 6388 & $18.90$ & $6.95$ & $1.54$ & $15.38$ & $0.24$ & $0.38$&$-0.13\pm0.07$&R05 \\            
NGC 6397 & $3.48$ & $6.06$ & $2.28$ & $11.89$ & $0.29$ & $0.18$&$-0.37\pm0.11$& F99\\             
NGC 6441 & $17.98$ & $7.35$ & $1.50$ & $15.82$ & $0.26$ & $0.45$&$-0.02\pm0.12$&R05 \\            
NGC 6522 & $6.70$ & $8.34$ & $1.71$ & $14.77$ & $0.28$ & $0.50$& &R05 \\            
NGC 6535 & $2.15$ & $10.53$ & $1.66$ & $14.29$ & $0.45$ & $0.32$&$-0.50\pm0.18$& F99\\            
NGC 6624 & $5.54$ & $8.00$ & $1.66$ & $14.45$ & $0.21$ & $0.27$&$-0.32\pm0.16$&R05 \\             
NGC 6681 & $8.21$ & $8.12$ & $1.68$ & $14.99$ & $0.10$ & $0.07$&$-0.82\pm0.09$&R05 F99\\          
NGC 6712 & $4.03$ & $8.03$ & $2.03$ & $14.15$ & $0.15$ & $0.46$&$0.02\pm0.05$ &R05 F99\\          
NGC 6752 & $4.50$ & $5.75$ & $2.06$ & $13.13$ & $0.45$ & $0.04$&$-0.03\pm0.15$ & F99\\             
NGC 6809 & $4.12$ & $6.96$ & $2.21$ & $13.78$ & $0.29$ & $0.07$& & F99\\            
NGC 6864 & $10.30$ & $8.51$ & $1.46$ & $16.63$ & $0.18$ & $0.16$& &R05 \\           
NGC 6934 & $4.91$ & $8.86$ & $1.62$ & $16.03$ & $0.11$ & $0.11$& &R05 F99\\         
NGC 7078 & $12.98$ & $6.42$ & $1.78$ & $15.19$ & $0.11$ & $0.09$&$-0.66\pm0.11$&R05 F99\\         
NGC 7089 & $7.39$ & $6.46$ & $1.85$ & $15.35$ & $0.15$ & $0.05$&$0.05\pm0.11$&R05 \\             
NGC 7099 & $5.13$ & $7.48$ & $1.93$ & $14.72$ & $0.08$ & $0.03$&$-0.57\pm0.11$&R05 F99\\          
\end{longtable}
\noindent$^a$ km/s, \cite{1993ASPC...50..357P}\\
$^b$ V-band mag, this paper\\
$^c$ arcsec, this paper\\
$^d$ mag, \cite{1999AJ....118.1738F} and \cite{2005A&A...432..851R}\\
$^e$ \cite{1999AJ....118.1738F, 2005A&A...432..851R} and this paper (see text)\\
$^f$ mag, \cite{2005A&A...432..851R} and \cite{1996AJ....112.1487H}\\
$^g$ dimensionless, \cite{2006AJ....132..447N}\\
$^h$ F99 indicates that the GC is listed by \cite{1999AJ....118.1738F}, R05 that it is listed by \cite{2005A&A...432..851R}.\\
}
\end{document}